\documentstyle{article}
\begin{document}

\null
\vspace{2cm}
\begin{center}
\begin{Large}
\begin{bf}
On the measurement of the Hubble constant \\
in a local low-density universe\\
\end{bf}
\end{Large}

\bigskip
\bigskip
\bigskip
\bigskip

Xiang-Ping Wu$^1$, Zugan Deng$^{1,2}$, Zhenlong Zou$^1$, Li-Zhi Fang$^{3}$
\& Bo Qin$^1$\\

\bigskip

{\noindent}$^1$ Beijing Astronomical Observatory,
         Chinese Academy of Sciences, Beijing 100080, China\\
{\noindent}$^2$ Department of Physics, Graduate School,
         Chinese Academy of Sciences, Beijing 100039, China\\
{\noindent}$^{3}$ Department of Physics and Steward Observatory,
	 University of Arizona, Tucson, AZ 85721, USA\\
\bigskip
\bigskip
\bigskip
\bigskip
\bigskip
\bigskip
\bigskip
\bigskip
\bigskip

\begin{large}
Submitted to\\

\bigskip
\bigskip

{\bf The Astrophysical Journal Letters} \\

\bigskip

{\it (Received ~~~~~~~~~~~~~)}\\

\bigskip
\bigskip
\bigskip

\end{large}

\bigskip
\bigskip
\bigskip
{\noindent}--------------------------------------------------------------------------------------------\\
{\noindent}E-mail: wxp@bao01.bao.ac.cn ~~ also ~~ wxp\%melamb@mesiob.obspm.fr
\end{center}

\newpage

\begin{Large}
\begin{bf}
\begin{center}
{\noindent}ON THE MEASUREMENT OF THE HUBBLE CONSTANT\\
IN A LOCAL LOW-DENSITY UNIVERSE\\
\end{center}
\end{bf}
\end{Large}

\begin{large}
\begin{bf}
\begin{center}
{\noindent}Xiang-Ping Wu$^1$, Zugan Deng$^{1,2}$, Zhenlong Zou$^1$,
Li-Zhi Fang$^{3}$ \& Bo Qin$^4$\\
\end{center}
\end{bf}
\end{large}

{\noindent}$^1$ Beijing Astronomical Observatory,
         Chinese Academy of Sciences, Beijing 100080, China\\
{\noindent}$^{2}$ Department of Physics, Graduate School,
         Chinese Academy of Sciences, Beijing 100039, China\\
{\noindent}$^{3}$ Department of Physics and Steward Observatory,
	 University of Arizona, Tucson, AZ 85721, USA\\


\bigskip

\begin{large}
\begin{center}
ABSTRACT\\
\end{center}
\end{large}

Astrophysical observations indicate that the ``Local Universe" has
a relatively lower matter density ($\Omega_0$) than
the predictions of the standard inflation cosmology and the large-scale
motions of galaxies which provide a mean mass density
to be very close to unity.   In such a local underdense region the
Hubble expansion may not be representative of the global behaviour.
Utilizing an underdense sphere embedded in
a flat universe as the model of our ``Local Universe",
we show that the local Hubble constant would be 1.2 -- 1.4
times larger than the global value on scale of $\sim80$ Mpc,
depending on the variation of $\Omega_0$. This may account for the recent
measurements of the unpleasantly large Hubble constant of
$\sim$80 km/s/Mpc using the Cepheid variables in the Virgo cluster
and the relative distance between Virgo and Coma cluster
and removes the resulted apparent paradox of the age of our universe. \\

{\noindent}{\it Subject headings:} cosmology: distance scale --
                                  large-scale structure of universe \\

\bigskip
\bigskip
\bigskip

\begin{large}

\begin{center}
1. INTRODUCTION\\
\end{center}

It is firmly established from the dynamical analysis that the mean matter
density of the Local Supercluster on scale of $\sim10$ Mpc is within a
factor of two of $\Omega_0=0.1$,  which is about an order of magnitude smaller
than the global value of $\Omega_0=1$ favoured by the inflation cosmological
model. Indeed, the large-scale motions of galaxies with the assumption of
``light-trace-mass" suggest an increasing tendency of $\Omega_0$ with
scale up to $\sim100$ Mpc [see Dekel (1994) for a recent review].
It is likely that we are situated
in a local low-density ``universe".   Therefore, the local properties can
significantly differ from the global ones and any local astrophysical
measurements may not actually be representative of the real universe.
Although this general argument is not entirely a new one [Readers are
recommended to refer to Turner, Cen \& Ostriker (1992, references therein)],
not much attention has been
really concentrated on the issue, due to the unknown details of theoretical
consideration on how large the differences between the local measurements and
the global ones would be. \\

The critical question arises from the two recent independent measurements of
the Hubble constant through the observations of Cepheid variables in the Virgo
cluster (Pierce et al. 1994; Freedman et al. 1994),
both of which give rise to a high value of $H_0\approx80$
km/s/Mpc. In the frame of standard cosmological model that assumes a completely
homogeneous matter distribution everywhere in a flat universe, this large
$H_0$ leads to a conflict that the expansion age of the Universe ($\sim8$ Gyr)
is smaller than the age of globular clusters of the Galaxy ($\sim16$ Gyr),
implying that either the standard cosmological model needs to be revised or
the present measurements of stellar ages need to be re-examined.
This situation is indeed unfortunate  as the standard Big Bang model and the
theories of stellar evolution are two fundamentals of modern astrophysics and
any modifications may have significant impacts on our conventional views.
To avoid the difficulties resulted from the new measurements of $H_0$,
we explore the possibility of attributing the high $H_0$ to a local
low-density region which expands faster than the rest of the universe.
This may survive the standard cosmological and stellar evolutionary theories.
\\

\bigskip

\begin{center}
2. MODEL AND RESULTS\\
\end{center}

The Tolman-Bondi metric is often used for the description of the space-time
of an overdense or an underdense spherical region embedded in an expanding
Universe (e.g. Zel'dovich \& Grishchuk 1984; Arnau et al. 1993; Fang \& Wu
1993;
Wu \& Fang 1994):
\begin{equation}
ds^2=\frac{r'^2}{1-\epsilon f^2(x)}dx^2+
     r^2(x,t)(d\theta^2+\sin^2\theta d\phi^2)-dt^2.
\end{equation}
The evolution of $r(x,t)$ is given by
\begin{equation}
\dot{r}^2=-\epsilon f^2(x)+\frac{F(x)}{r}
\end{equation}
and $F(x)$ relates with the invariant mass $M(x)$ within $x$ through
\begin{equation}
F(x)=2M(x)=\int_0^r 8\pi x^2\rho(x,t) dx \equiv \frac{8\pi}{3}r_i^3
\overline{\rho}_i
\end{equation}
where the prime denotes a derivative with respect to $x$, the dot with
respect to $t$, $\rho$ and $\overline{\rho}$ are the matter density and the
mean
matter density, respectively.  We use the subscript ``$_i$" to stand for the
``initial epoch" and ``$_o$" for the ``present time".   \\

Assuming a flat universe as the background, we can write the solution to
an underdense region as ($\epsilon=-1$)
\begin{Large}
\begin{equation}
\begin{array}{l}
\frac{S}{S_i}=\frac{1-\delta_i}{\delta_i}\frac{\cosh\eta-1}{2}\\
\sinh\eta-\eta=\frac{4}{3}\frac{\delta_i^{3/2}}{1-\delta_i}(\frac{t}{t_i}-1)
               +\frac{2\delta_i^{1/2}}{1-\delta_i}
               -\cosh^{-1}\frac{1+\delta_i}{1-\delta_i}.
\end{array}
\end{equation}
\end{Large}
{\noindent}Here $S$ is the expansion factor
of the universe defined by $r=S(x,t)x$, $\delta_i$, the mean initial matter
perturbation given by $\overline{\rho_i}=\rho_{ci}(1-\delta_i)$, and $\rho_c$,
the critical mass density of the universe. The present Hubble constant
is introduced through $H_0=\dot{S_0}/S_0$, which is generally a function of
the radial coordinate $x$, i.e., the expansion rate may vary not only
with time but also with position.  We obtain the relation
between the present local Hubble constant ($H_L$) in the underdense region and
the present global Hubble constant ($H_G$) in the background universe to be
\begin{equation}
\frac{H_L}{H_G}=\frac{(1+z_i)^{3/2}}{(S_0/S_i)}\left[\delta_i+
                \frac{1-\delta_i}{(S_0/S_i)}\right]^{1/2}
\end{equation}
in which $z_i$ is the redshift at the epoch $t=t_i$ when the
initial density perturbation occurred.  For simplicity,
we choose $t_i$ to be the decoupling
time that corresponds to $z_i\approx1000$. The present density contrast
$\frac{\Delta\rho}{\rho}$ stemming from the initial density perturbation
profile
$\delta(x)$ is found to be
\begin{equation}
\frac{\Delta\rho}{\rho}=1-\left[\frac{1+z_i}{(S_0/S_i)}\right]^3
                        \frac{1-\delta(x)}{1+d\ln S_0/d\ln x}
\end{equation}
and the present density parameter is $\Omega_0=1-\frac{\Delta\rho}{\rho}$. \\

We first adopt a constant density perturbation profile
$\delta(x)=\delta_i=\delta_0$ in the local underdense region.
In this case, the present Hubble constant
inside the underdense sphere depends uniquely on the present matter density
$\Omega_0$, as is shown in Figure 1. In fact, this corresponds to an ensemble
of solutions to the expanding ``universe" with different $\Omega_0(<1)$.
Utilizing this model to our local universe, a present  matter density of
$\Omega_0=0.2$ results in a local Hubble constant of 1.33 times larger than
the global one and the $\Omega_0=0.5$ gives $H_L=1.19H_G$. In the extreme
case of $\Omega_0=0$, the local expansion rate is 1.5 times larger than that
of the background universe of $\Omega_0=1$. Therefore, if we are unfortunately
situated in a local low-density universe with a mass density of a few tenth
of the critical value of the background universe, the local measurements
like the two recent observations (Pierce et al. 1994; Freedman et al. 1994)
utilizing the Cepheid variables of the Virgo cluster and the recession
velocity of the Coma cluster would provide a relatively higher Hubble
constant than the true value in the background flat universe.\\

Nevertheless, the local universe on scale of as large as the distance to
Coma cluster cannot be well described by a constant matter perturbation.
It appears that $\Omega_0$ varies from $\sim0$ to $\sim1$ with the increase
of scale.  We have then tested two initial density perturbation profiles that
give rise to the similar shape of $\Omega_0$ to the observed one:
$\delta(x)=\delta_0/[1+(x/a)^2]$ (the isothermal sphere with a core)
and $\delta(x)=\delta_0/[1+(x/a)^2]^{3/2}$ (the King model),
where $\delta_0$ determines the maximum initial density contrast and
$a$ ($a_0$) is the initial (present) scale length of the  perturbed region.
Our computations show that these two profiles don't provide significantly
different results of the present density contrast within $\sim100$ Mpc.
Figure 2 demonstrates the variations of $\Delta\rho/\rho$ with distance
for three sets of parameters in the King model. Although these curves
might not exactly fit to the true distribution of $\Omega_0$
which has been unknown to date, they
essentially represent the variation tendency of $\Omega_0$ with scale,
which provide $\Omega_0\sim0.1$ on scale of 10 Mpc and
$\Omega_0\sim(0.4$ -- $0.9$) on scale of 100 Mpc. The Hubble constant
variations with distance are shown in Figure 3 for the same parameters
in Figure 2.  At the distance of $\sim80$ Mpc where the Coma cluster
locates as were indicated by the recent observations, the local Hubble
constant may be estimated to be 1.2 -- 1.4 times larger than the global
one, depending on the local matter content. \\

\bigskip

\begin{center}
3. DISCUSSION\\
\end{center}

Recent measurements of the Hubble constant using the Cepheid variables
in the Virgo cluster and the relative distance between the Virgo and the
Coma cluster result in $H_0=87\pm7$ km/s/Mpc (Pierce et al. 1994) and
$H_0=80\pm17$ km/s/Mpc (Freedman et al. 1994),
which disagree with the other two HST measurements (Sandage et al. 1994;
Saha et al. 1994) of $H_0=52\pm9$ km/s/Mpc
using the Cepheid variables and the brightness of the type Ia supernovae
in two relatively closer galaxies (distance=4.1 -- 4.7 Mpc).  The former
determines actually the expansion rate of the Coma cluster which is so
distant that the peculiar velocity contributes a negligible component and
furthermore, the different methods produce the same mean relative
Coma-Virgo distance modules.
Whilst the later may suffer from the local calibration of the brightness of
type Ia supernovae, leading to an overestimate of distance (Hogan 1994).
We believe
that the Coma recession velocity and the Coma-Virgo relative distance
measurements are likely to reflect the nature of the local universe.  \\

The apparent paradox of the ``young" age of the universe may have arisen
from the
misuse of the local Hubble constant $H_L$ as the global value $H_G$. The
Hubble constant of the background universe can be estimated by
reducing the measured Hubble constant by a factor of $1.2$ -- $1.4$ at
the distance of Coma cluster, leading to an increase of the currently
estimated age of the universe by the same factor. Thus, the cosmological
conflict between the expansion age of the universe, predicted in the standard
cosmological model using the recent measurements of the large Hubble
constant of $H_0\sim80$ km/s/Mpc, and the ages of the oldest globular
clusters of the Galaxy may vanish, or at least is partially resolved.\\

It appears that the true Hubble constant of the universe can be directly
measured only when the observations are made beyond the local low-density
region.  It then remains to be promising that the time delay between the
images of gravitationally lensed quasars at the redshift of $\sim1$ may provide
the reliable value of $H_0$. From the optical/radio monitoring of the double
quasar 0957+561A,B ($z=1.41$) over $\sim10$-years coverage (Vanderriest
et al. 1989; Roberts et al. 1991), which exhibits a
time delay of 415/513 days, and the theoretical modeling of the lensing
galaxies, one finds a Hubble constant of $H_0=48^{+16}_{-7}$ /$39^{+13}_{-6}$.
Another evidence for supporting a low value of $H_0\sim50$ km/s/Mpc obtained
in the distant universe comes from the measurement of the Sunyaev-Zel'dovich
effect in Abell cluster A2218 ($z=0.171$) (Jones et al. 1993).
Indeed, these two measurements of
the Hubble constant at the cosmological distance yield a value apparently
smaller than the local one, indicative of an expansion age of the
universe comparable with the age of the oldest globular clusters.
Bartlett et al (1994) have even claimed for a Hubble constant of as small as
$30$ km/s/Mpc  and found that the small $H_0$ can overcome
most of the difficulties in the current standard cosmological model.
Further measurements of dynamical properties of the local universe
and the future HST observations of the Cepheid variables in  more distant
galaxies are needed to confirm our arguments.\\

We thank Gary Mamon for reading the manuscript and for valuable comments.
This work has been supported by the China National Science Foundation
and the Chinese Academy of Sciences. \\


\begin{large}
\begin{center}
References\\
\end{center}
\end{large}

{\noindent}Arnau, J. V., Fullana, M. J., Monreal, L., \& S\'aez, D.
           1993, ApJ, 402, 359\\
Bartlett, J. G., Blanchard, A., Silk, J., \& Turner, M. S., 1994, Science,
             in press\\
Dekel, A. 1994, ARA\&A, 32, 371.\\
Freedman, W. L., et al. 1994, Nature, 371, 757\\
Fang, L. Z., \& Wu, X. P. 1993, ApJ, 408, 25\\
Hogan, C. J. 1994, Nature, 371, 374\\
Jones, M., et al. 1993, Nature, 365, 320\\
Pierce, M. J., et al. 1994, Nature, 371, 385\\
Roberts, D. H., Leh\'ar, J., Hewitt, J. N., \& Burke, B. F. 1991,
        Nature, 352, 43\\
Saha, A., et al. 1994, ApJ, 425, 14\\
Sandage, A. et al. 1994, ApJ, 423, L13\\
Turner, E. L., Cen R., \& Ostriker, J. P. 1992, AJ, 103, 1427\\
Vanderriest, C., et al. 1989, A\&A, 215, 1\\
Wu, X. P., \& Fang, L. Z. 1994, ApJ, 424, 530\\
Zel'dovich, Ya. B., \& Grishchuk, L. P. 1984, MNRAS, 207, 23p\\

\end{large}

\newpage
\begin{large}
\begin{center}
Figure Captions\\
\end{center}

\bigskip
\bigskip

{\noindent}{\it Figure 1} ~~ The ratio of the local Hubble constant $H_L$
to the global value $H_G$ versus $\Omega_0$. The local underdense region
is modelled by a sphere with constant matter density $\Omega_0$ embedded
in a flat universe.\\

\bigskip
\bigskip

{\noindent}{\it Figure 2} ~~ Variations of local matter density $\Omega_0$
with scale for the King model as the local negative density perturbations.
Two parameters determine the model: the maximum initial density fluctuation
$\delta_0$ and the present length scale $a_0$. \\

\bigskip
\bigskip

{\noindent}{\it Figure 3} ~~ Variations of $H_L/H_G$ with scale for the
three sets of parameters in Figure 2.

\end{large}
\end{document}